\documentclass[aps, prb, twocolumn,amsmath,amssymb,showpacs,superscriptaddress]{revtex4-1}
\usepackage{graphicx}
\usepackage{comment}
\begin{document}

\preprint{}


\renewcommand{\vec}[1]{\mathbf{#1}}

\title{Lattice dynamics in the paraelectric phase of PbHfO$_3$ studied by inelastic X-ray scattering}

\author{R. G. Burkovsky}
\email{burkovsk@esrf.fr}
\affiliation{European Synchrotron Radiation Facility, 71 avenue des Martyrs, 38000 Grenoble, France}
\affiliation{St.Petersburg State Polytechnical University, 29 Politekhnicheskaya, 195251, St.-Petersburg, Russia}

\author{D. Andronikova}
\affiliation{St.Petersburg State Polytechnical University, 29 Politekhnicheskaya, 195251, St.-Petersburg, Russia}

\author{Yu. Bronwald}
\affiliation{Ioffe Phys.-Tech. Institute, 26 Politekhnicheskaya, 194021, St.-Petersburg, Russia}


\author{M. Krisch}
\affiliation{ European Synchrotron Radiation Facility, 71 avenue des Martyrs, 38000 Grenoble, France }


\author{K. Roleder}
\affiliation{Institute of Physics, University of Silesia, ul. Uniwersytecka 4, 40-007 Katowice, Poland}

\author{A. Majchrowski}
\affiliation{Institute of Applied Physics, Military University of Technology, ul. Kaliskiego 2, 00-908 Warsaw, Poland}

\author{A. V. Filimonov}
\affiliation{St.Petersburg State Polytechnical University, 29 Politekhnicheskaya, 195251, St.-Petersburg, Russia}

\author{A. I. Rudskoy}
\affiliation{St.Petersburg State Polytechnical University, 29 Politekhnicheskaya, 195251, St.-Petersburg, Russia}

\author{S. B. Vakhrushev}
\affiliation{Ioffe Phys.-Tech. Institute, 26 Politekhnicheskaya, 194021, St.-Petersburg, Russia}

\date{\today}

\begin{abstract}

We report the results of  an inelastic X-ray scattering study of the lattice dynamics in the paraelectric phase of the antiferroelectric lead hafnate PbHfO$_3$. The study reveals an avoided crossing between the transverse acoustic and transverse optic phonon modes propagating along the [1 1 0] direction with [1 -1 0] polarization. The static susceptibility with respect to the generally incommensurate modulations is shown to increase on cooling for the entire $\Gamma$-M direction. We consider different approaches to the data analysis that correspond to different models for the temperature evolution of the dynamic susceptibility function. 
A number of similarities and differences between the lattice dynamics of PbHfO$_3$ and PbZrO$_3$ are described. 

\end{abstract}



\maketitle

\section{Introduction}

Perovskite ferroelectrics and similar crystals find numerous applications in technology \cite{Scott2007}. These materials form a rich test ground for assessing the phenomena associated with the structural phase transitions. Since the introduction of the soft mode concept by Cochran \cite{Cochran1960} and Anderson \cite{Anderson1960} in the context of the ferroelectric transition in BaTiO$_3$, the topic of the lattice dynamics in perovskites attracted considerable attention. A number of thorough reviews were published before 1990 \cite{Scott1974,Blinc1974,Lines1979,Bruce1981} that considered the results of the soft mode spectroscopy in various perovskites.
More recently an inelastic X-ray scattering (IXS) study of the critical dynamics in PbZrO$_3$-like antiferroelectrics revealed that the critical dynamics in the paraelectric phase of PbZrO$_3$ is compatible with the one expected for the crystal with an incommensurate phase transition \cite{Tagantsev2013,Burkovsky2014}. The key feature is a flat transverse acoustic (TA) branch, which shows an almost homogeneous softening along a large part of the $\Gamma$-M (wavevector $\vec{q}=(q,q,0)$) direction \cite{Tagantsev2013,Burkovsky2014}. 

In this paper we report the IXS study of a similar crystal, lead hafnate, PbHfO$_3$ \cite{Shirane1953}. It is different from PbZrO$_3$ by the presence of an intermediate phase between the cubic phase and low-temperature antiferroelectric phase \cite{Shirane1953,Dernier1975,Leontyev1984,Fujishita2002,Kupriyanov2012}. The structure of the intermediate phase (or phases, according to some reports \cite{Kupriyanov2012}) is currently under debate. The current study covers the lattice dynamics in the high-temperature cubic phase.


In section \ref{sec_methods} we specify the experimental methods, including sample preparation and IXS measurements. In section \ref{sec_spectra_modeling} we specify the basic model used to fit the IXS spectra. 
The experimental results are covered in sections \ref{sec_results_dispersion} and \ref{sec_results_evolution}, dedicated respectively to the analysis of dispersion curves at high temperatures and to the analysis of the temperature evolution of the signal.
A compact summary of the results and the data treatment is given in section \ref{sec_summary}. The last two sections are dedicated to the analysis of possible interpretations and for the conclusion.

\section{Experimental methods} \label{sec_methods}
\subsection{Sample}
The single crystals of PbHfO$_3$ were synthesized by means of spontaneous crystallization from the high temperature solution in a Pb$_3$O$_4$-B$_2$O$_3$ solvent. The composition of the melt used in our experiments was as follows: 2.4 mol\% of PbHfO$_3$, 77 mol\% of PbO (re-counted to Pb$_3$O$_4$) and 20.6 mol\% of B$_2$O$_3$. The Pb$_3$O$_4$ was used instead of PbO to avoid coloration of the as-grown crystals caused by oxygen deficiency. The crystallization was carried out in a platinum crucible covered with a platinum lid under conditions of a low temperature gradient. After soaking at 1473 K for 24 hours the melt was cooled down to 1200 K at a rate of 3.5 K/h and after decantation the furnace was cooled to room temperature at a rate of 10 K/h. As-grown PbHfO$_3$ single crystals were etched in diluted acetic acid to remove residues of the solidified flux. 
The sample for the IXS measurements was a 40 by 40 by 600 micron size needle. 
In accord with the literature, the cell parameter of the cubic phase was found to be $a=4.14$ at $T=473$ K.

\subsection{IXS measurements}
The experiment was performed at ID28 beamline of the ESRF (European Synchrotron Radiation Facility), using the Si (9,9,9) setup at $E=17.794$ keV which provides an overall energy resolution of about 3 meV \cite{Krisch2007}. We studied the phonons propagating along the [1 0 0] and [1 1 0] direction with polarization (direction of the atomic displacements) lying in the plane defined by these two vectors. Thanks to the fact that the spectrometer hosts 9 independent analyzer-detector pairs, the spectra at points away from these directions but within the corresponding plane were also recorded. The sample was glued to quartz capillary. The temperature was controlled by the ESRF mini heat blower from 473 K (just at the bottom of the stability region of the cubic phase) to 773 K with an accuracy of about 2 K.

\section{IXS spectra modeling} \label{sec_spectra_modeling}
The modeling of the spectra was done following the standard approach for which the one-phonon scattering function for a phonon with wavevector $\vec{q}$ in branch $j$ is  \cite{Dorner1982}
\begin{equation} \label{eq_sc_fun}
S_j(\vec{Q},\omega) = |G_j (\vec{Q},\vec{q}) |^2 F_j\left[\omega,\omega_j(\vec{q}),T\right],
\end{equation}
where $G_j (\vec{Q},\vec{q})$ -- the inelastic structure factor and $F_j[\omega,\omega_j(\vec{q}),T]$ -- the response function, $\hbar \omega$ -- the energy transfer in the one-phonon scattering event and $\omega_j(\vec{q})$ describes the dispersion curve. The magnitudes of the inelastic structure factors \cite{Dorner1982} were treated in our analysis as variable parameters. The model response function corresponds to the dynamic susceptibility of a damped harmonic oscillator, given by \cite{Dorner1982}
\begin{equation} \label{eq_chi_dho}
\chi^\text{Dyn}_\text{DHO} = (\omega_j^2-\omega^2-i\omega\Gamma)^{-1},
\end{equation}
where $\Gamma$ is a damping constant. The resulting expression for the response function is \cite{Dorner1982}
\begin{eqnarray} \label{eq_rf}
F_j[\omega,\omega_j(\vec{q}),T] = \nonumber \\
\frac{\omega}{1-\text{exp}(-\hbar\omega/kT)}
\frac{\Gamma_j(\vec{q},T)}{[\omega^2-\omega^2_j(\vec{q},T)]^2 + \omega^2 \Gamma^2_j(\vec{q},T)}.
\end{eqnarray}

The IXS signal was modeled by a numerical convolution of the sum of the scattering functions \eqref{eq_sc_fun} (DHO) with the resolution function of the IXS spectrometer. At a number of occasions we also added the resolution-limited central peak (CP). The resolution function was modeled by a pseudo-voigt lineshape with independently determined parameters.

\section{Results: phonon dispersion at high temperature} \label{sec_results_dispersion}
\begin{figure}
\includegraphics[width=.99\columnwidth, clip=false, trim= 0 0 0 0]{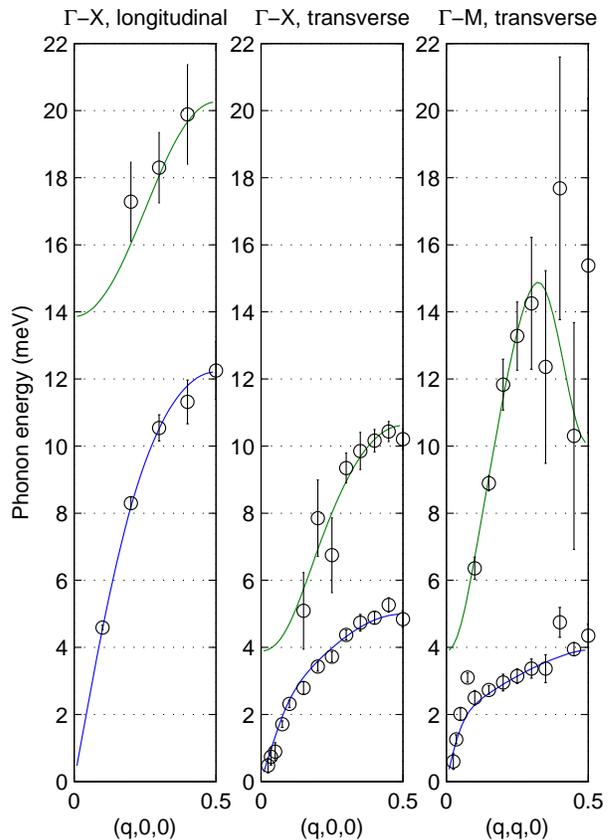}
\caption{\label{fig_ixs}
Phonon dispersion curves of PbHfO$_3$ at T=773 K. The solid lines show a fit by the rigid ion model.
}
\end{figure}

We start from the general overview of the dispersion curves at high temperatures. At T=773 K the two phonon resonances are well resolved. This allows us postponing the discussion of the less straightforward low temperature data treatment process to the next sections. The dispersion curves are shown in Fig. \ref{fig_ixs}. 

The key feature is the TA (transverse acoustic) branch, which displays a relatively weak $q$-dispersion in $\Gamma$-M direction ($\vec{q}=(q,q,0)$) and appears relatively flat for $q>0.075$ (hereafter $q$ is expressed in reciprocal lattice units, r.l.u.). This is similar to the result of Refs. \onlinecite{Tagantsev2013,Burkovsky2014} for PbZrO$_3$. Approximately at $q=q_{AC}\approx 0.075$ there is a rapid change in the group velocity. The linear slope of the transverse optic (TO) branch at $q>q_{AC}$ appears to be approximately the same as the linear slope of the TA branch for $q<q_{AC}$ (which corresponds to the transverse speed of sound). So, they can be viewed as continuations of each other. As it appears from Fig. \ref{fig_ixs} there is an avoided crossing (also referred to as anti-crossing) between the TA and TO branches. In perovskites this phenomenon is not uncommon. In particular the studies of KNbO$_3$ \cite{Fontana1981} and BaTiO$_3$ \cite{Shirane1967} show a similar behavior. The shell model fit in KNbO$_3$ suggests \cite{Fontana1981} that the flat part of the TA branch largely corresponds to the motion of the ferroelectric-active niobium ion. By analogy one may suggest that the motion in the flat part of the TA branch corresponds to the motion of the ferroelectric-active Pb$^{2+}$ ion. A similar assumption was made in formulating the simple model (``polarizability model'') for PbZrO$_3$ and PbHfO$_3$ in Ref. \cite{Bussmann2014}.

We performed the analysis of the motion in the simplest possible harmonic model –- the rigid ion model (RIM). For the perovskite structure this model is described in Refs.  \onlinecite{Cowley1964,Stirling1972}. The solid lines in Fig. \ref{fig_ixs} corresponds to the least squares fit by RIM with the following parameters: longitudinal force constants A$_1$=20.15, A$_2$=289.1, A$_3$=-11.22, transverse force constants B$_1$=-3.944, B$_2$=-63.09, B$_3$=3.595 (in units $e^2/2v_a$ where $e$ -- electron charge and $v_a$=(4.15 \AA)$^3$ -- unit cell volume), charges of lead and hafnium $Z_1=1.7e$ and $Z_2=4.07e$. The fit by RIM indicates that the TA branch in the $\Gamma$-M direction, for $q>q_{AC}$, corresponds mainly to the motion of lead and hafnium with displacements (not normalized by the square root of the mass) of lead being about 2 times larger than the ones of hafnium. This supports the assumption of mainly lead motions for the flat TA branch. In principle, a more reliable analysis could be possible with more detailed harmonic models, such as the shell model, which accounts for the electronic polarizability of the ions. This type of analysis is out of scope of the current paper, in part due to the absence of reliable data on the high-energy modes in the paraelectric phase of lead hafnate, which are required for such analysis \cite{Cowley1964}.

\section{Results: temperature evolution of the IXS signal} \label{sec_results_evolution}

\subsection{Relations for susceptibilities}
The inelastic and quasielastic scattering in the high-symmetry phase is used in this study to gather the information on the susceptibility of the crystal with respect to various sorts of ionic displacements that break the high symmetry \cite{Bruce1981}. It is often the case that on approaching the structural phase transition of a particular type, the static susceptibility with respect to the corresponding patterns of ionic displacements grows and even diverges at the transition point. The simplest case, which we assume throughout the paper, is when the transition is second order like and the static susceptibility obeys the Curie-Weiss (CW) law 
\begin{equation} \label{eq_cw}
\chi^\text{Stat}(T) = \frac{C}{T-T_0},
\end{equation}
where $C$ is the CW constant.
The static susceptibility in this formula should not be confused with the dynamic susceptibility. The connection between these two susceptibilities is given by the Kramers-Kronig relation \cite{Bruce1981}
\begin{equation} \label{eq_kk}
\chi^\text{Stat} = \text{Re}\left[\chi^\text{Dyn}(\omega=0)\right] = 
\frac{1}{\pi} \int_{-\infty}^{\infty} \frac{\text{Im}\left[\chi^\text{Dyn}(\omega')\right]}{\omega'} d\omega'
\end{equation}

The static one-phonon susceptibility is defined by the matrix \cite{Bruce1981}
\begin{equation} \label{eq_chi_jj}
\chi^\text{Stat}_{jj'}(\vec{q})=\beta\langle Q_j(\vec{q}) Q_{j'}(\vec{-q}) \rangle,
\end{equation}
where $\beta=(kT)^{-1}$, $\langle\rangle$ -- ensemble average and $Q_j(\vec{q})$ -- normal mode coordinates. In the case of susceptibility with respect to the ionic displacements within a particular phonon mode, the diagonal terms in Eq. \eqref{eq_chi_jj} are given (in the high-temperature approximation) by \cite{Bruce1981}
\begin{equation} \label{eq_chi_omega}
\chi^\text{Stat}_j(\vec{q})=\frac{1}{\omega^2_j(\vec{q})}.
\end{equation}
On the other hand, this static susceptibility is proportional to the integral of the scattering function \eqref{eq_sc_fun} over all energy transfers
\begin{equation} \label{eq_sc_fun_int}
\chi^\text{Stat}_j(\vec{q}) = \frac{\beta}{|G_j (\vec{Q},\vec{q}) |^2} S^\text{Int}_j(\vec{Q}).
\end{equation}
Finally, as it might be the case of PbHfO$_3$, the crystal may have degrees of freedom that are not of harmonic nature. In particular, they can be slowly-relaxing quasi-spin ionic shifts. In this case the dynamic susceptibility will not be of the form \eqref{eq_chi_dho}, but the relations \eqref{eq_kk},\eqref{eq_chi_jj} and \eqref{eq_sc_fun_int} will be still relevant, provided that $Q_j(\vec{q})$ and $G_j (\vec{Q},\vec{q})$ refer to the appropriately defined normal mode coordinates and structure factors. 

\subsection{Dielectric permittivity and the approach for analyzing the temperature dependences of finite-$q$ susceptibilities}

Due to the finite resolution effects, it is not possible to assess small-energy phonons at the Brillouin zone center by the IXS technique. Published studies of the temperature dependence of the dielectric permittivity suggested that $\chi^\text{Stat}_\text{Dielectric}$ obeys CW law. The CW temperature was reported as $T_0^\text{Shirane}= 323$ K by Shirane $et~al.$\cite{Shirane1953} and as $T_0^\text{Samara}= 378$ K by Samara\cite{Samara1970}. In these two studies the CW law was determined by analyzing the dependence of the dielectric permittivity $\epsilon(T)$ in the range $T_c < T <T_c+\Delta T$, where $\Delta T$ is about 90 K.
The recent report by Bussmann-Holder $et~al.$ \cite{Bussmann2015} (45 years after Samara's work), where $\epsilon(T)$ was measured in a broader temperature range, shows that the situation can be more complex. Although it appears that their data for $T_c < T <T_c+\Delta T$ are roughly consistent with $T_0=T_0^\text{Samara}= 378$ K (data from Ref. \onlinecite{Bussmann2015} would have given $T_0 = 404 $K assuming the same $T_c$ as in Samara’s paper), for higher temperatures, $T >T_c+\Delta$, the $\epsilon(T)$ obeys CW law with $T_0=T_0^\text{Bussmann}= 441$ K (In Ref. \onlinecite{Bussmann2015} $T_0$ is reported as 451 K, but the private communication with the authors indicated there was a misprint).
It is worth mentioning that $T_0^\text{Bussmann}= 441$ K is close to the temperature of the phase transition to the low-temperature antiferroelectric phase [$T_\text{A1-A2}$ in Ref. \onlinecite{Bussmann2015}].

In order to compare the temperature dependences of different kinds of susceptibility we define a function describing the relative rate of growth as
\begin{equation} \label{eq_rrog}
R(T) = -\frac{1}{\chi^\text{Stat}(T)}\frac{d\chi^\text{Stat}(T)}{dT}
\end{equation}
In terms of this function we may state that the relative rate of growth of $\epsilon(T)$, for $T_c < T <T_c+\Delta T$, is numerically similar to the one for $\chi^\text{Stat}(T)$ described by CW law with $T_0 \approx T_0^\text{Samara}$. The function $R(T)$ can be useful in comparing the temperature dependences of susceptibilities when the absolute scale of $\chi^\text{Stat}(T)$ is unknown, as it is the case in scattering experiments. In the case that the susceptibility can be described by $\chi^\text{Stat}=C/(T-T_0)^n$, the function $R(T)=n/(T-T_0)$ contains the information on both $T_0$ and power $n$.



\subsection{Modeling the temperature dependence of the signal at $\vec{q}=(0.1, 0.1, 0)$}

\begin{figure}
\includegraphics[width=.99\columnwidth, clip=false, trim= 10 0 10 0]{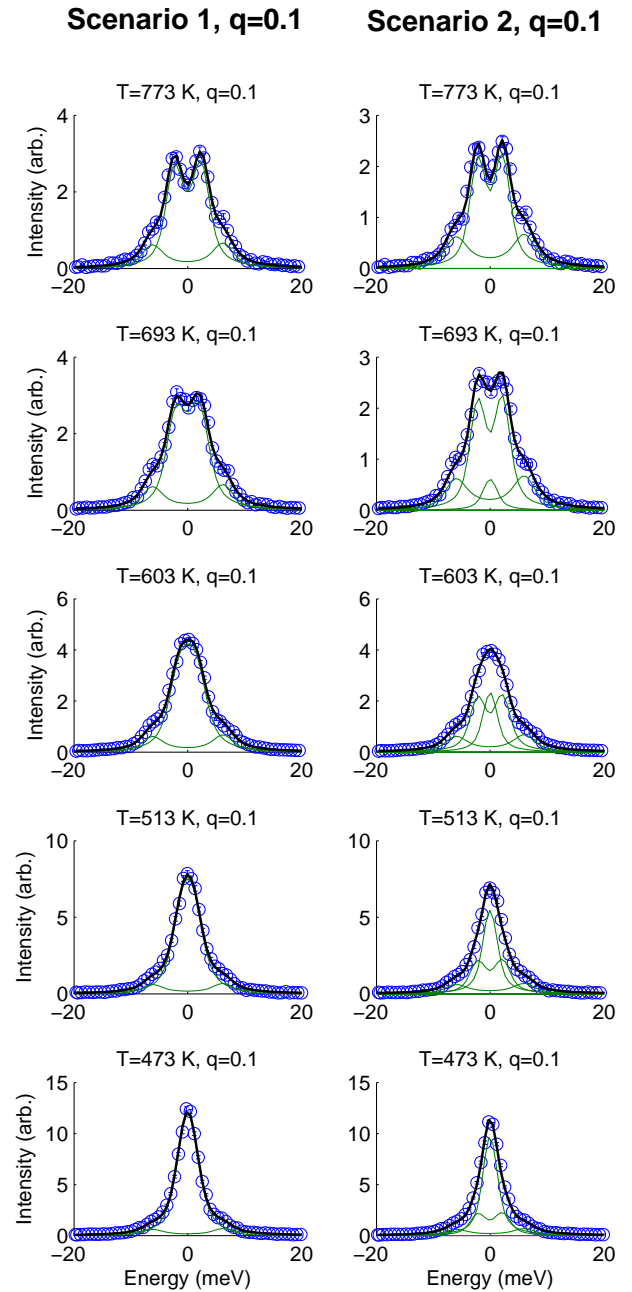}
\caption{\label{fig_q01_comparison}
Temperature evolution of the IXS spectra at $\vec{Q}=(2.1, 0.1, 0)$. The left and right panels show the same data fitted using the two different models described in the text. The data are normalized by the value $|G_\text{TO}|^2 T$ where $|G_\text{TO}|^2$ is the squared absolute value of the TO phonon structure factor, as obtained by the fit, and $T$ -- temperature.
}
\end{figure}

Fig. \ref{fig_q01_comparison} shows the temperature dependence of the signal at $\vec{q}=(0.1, 0.1, 0)$. The two panels (left and right) show the same data fitted by two different models that will be discussed in this section. It is evident that on temperature decrease the scattering at small energy transfers gets stronger. The overall shape of the low-energy part of the spectrum (everything excluding the second, high-energy phonon peak) gradually transforms from the underdamped, two-peak DHO lineshape to the single peak centered at zero energy transfer. At the lowest temperature the peak appears to be almost resolution limited. Neither the energy nor the damping constant of the second (TO) phonon appears to be temperature dependent, as deduced from visual inspection of the spectra and from the fits obtained by different methods that we employed. The data allows different approaches for their treatment. In this section we discuss 3 scenarios, in which we account for the temperature dependence of the signal by (1) temperature-dependent TA branch, (2) temperature-dependent central peak and (3) a combination of the two.

\subsubsection{Scenario 1. Temperature-dependent (soft) TA branch.}
This scenario is the formally simplest model function, consisting of two DHO contributions corresponding to the TA and TO phonons respectively. We set up the fitting process so that all the spectra, corresponding to a particular $q$ (currently: $q=0.1$) and the whole set of temperatures, are fitted together. This allows improving the reliability of the optimization problem by setting up constraints on the model parameters that are not possible with independent fitting of different spectra. We set up the following constraints:

\begin{itemize}
  \item The energy and the damping constant of the TO phonon is temperature-independent.
  \item The ratio between the structure factors of TA and TO phonons is temperature-independent.
\end{itemize}

The first constraint is motivated by the fact that we did not observe any notable difference in the relevant values when fitting the different spectra independently. The second constraint is due to the assumption that the structure factors do not change with temperature. The reason for this is that the pattern of the ionic displacements does not change with temperature and the temperature evolution manifests itself only in the temperature dependences of energy and damping constant of the phonon. In principle, the second constraint could be reformulated by fixing the absolute values for both structure factors. However, in this experiment we were unable to secure the same intensity scale for scans at different temperatures and different momentum transfers (this is generally challenging), so we were able to fix only their ratio. The result of the described fit is shown in the left panel of Fig. \ref{fig_q01_comparison}.

\begin{figure}
\includegraphics[width=.8\columnwidth, clip=false, trim= 0 0 0 0]{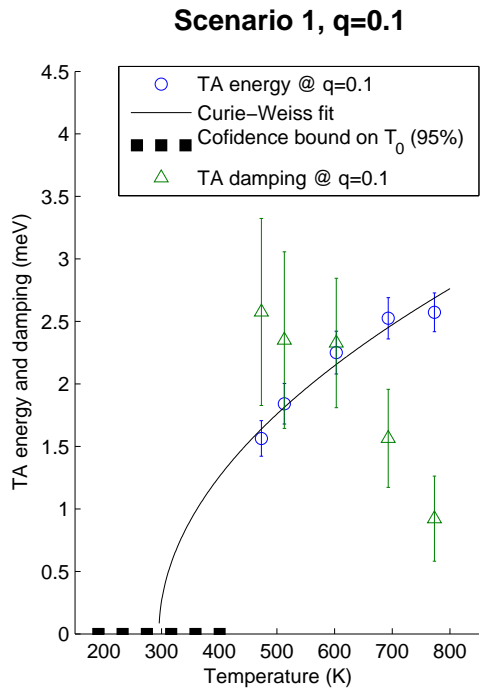}
\caption{\label{fig_nocp_cw}
Temperature dependence of the TA phonon energy and damping constant for $\vec{Q}=(2.1, 0.1, 0)$, obtained by the treatment in accord with the Scenario 1 described in the text. Error bars correspond to the 95\% confidence bounds. The Curie-Weiss fit to the phonon energy dependence is shown by solid line and the corresponding estimate of the Curie-Weiss temperature $T_0$ is shown by a dashed line.
}
\end{figure}

The temperature dependence of the TA phonon energy and of the damping constant are shown in Fig. \ref{fig_nocp_cw}. The energy gradually decreases while the damping gradually increases. The function Eq. \eqref{eq_rf} stops displaying the two-peak shape when $\Gamma_j > \omega_j \sqrt{2}$, which corresponds to overdamping of the phonon. From the dependences in Fig. \ref{fig_nocp_cw} it thus follows, that almost in the whole studied temperature region the response is underdamped and the apparent merging of the TA Stokes and anti-Stokes peaks already at $T=603$ K is a resolution effect. Overdamping is not excluded at the lowest studied temperature.

We obtain the model expression for the frequency of the TA phonon assuming that the corresponding susceptibility follows the CW law by combining Eqs. \eqref{eq_chi_omega} and \eqref{eq_cw}:

\begin{equation} \label{eq_sqrt}
\omega_\text{TA} = \sqrt{a(T-T^\text{TA}_0)  },
\end{equation}
where $a$ is a constant. Here we presume that the TA branch corresponds to the motion of the order parameter relevant to the phase transition characterized by modulation with a particular wavevector (in the current case $q=0.1$).

The corresponding least-squares fit is shown in Fig. \ref{fig_nocp_cw} as a solid line. This results in value $T^\text{TA}_0=295 \pm 115 K$. The corresponding relative rate of growth $R(T)$ (Eq. \eqref{eq_rrog}) is thus comparable to the one of $\epsilon(T)$.

\subsubsection{Scenario 2. Temperature-dependent central peak.}

In this scenario, in addition to the TA and TO phonon resonances, a resolution-limited central peak is present. The constraints are the following:

\begin{itemize}
  \item The energies and the damping constants of the TA and TO phonons are temperature-independent.
  \item The ratio between the structure factors of TA and TO phonons is temperature-independent (as in Scenario 1).
\end{itemize}

The fit by this model is shown in the right panel of Fig. \ref{fig_q01_comparison}. 
The simplest interpretation of the central peak is when we consider it as being due to an independent degree of freedom. The response of this degree of freedom to the external force is of relaxational character with a relaxation time (which determines its energy width) sufficiently longer than the period of the lattice vibrations. This relaxational degree of freedom may correspond to the hopping of ions between the wells of the multi-well local potential. We presume that this relaxational degree of freedom may be characterized by a susceptibility described by the CW law.
Combining equations \eqref{eq_cw} and \eqref{eq_sc_fun_int} the temperature dependence of the integrated central peak intensity can be written as
\begin{equation}\label{eq_icp_t}
S^\text{Int}_\text{CP}(T) = \frac{|G_j (\vec{Q},\vec{q}) |^2}{\beta} \frac{C}{T-T^\text{CP}_0}.
\end{equation}

\begin{figure}
\includegraphics[width=.8\columnwidth, clip=false, trim= 0 0 0 0]{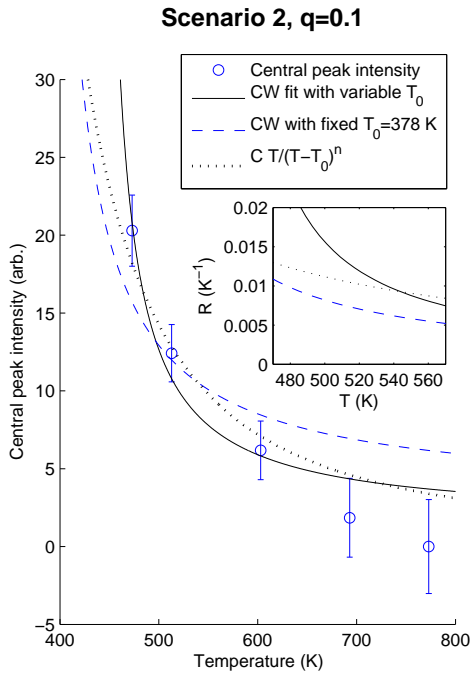}
\caption{\label{fig_cp_cw}
Temperature dependence of the central peak intensity for $\vec{Q}=(2.1, 0.1, 0)$, obtained by the treatment in accord with the Scenario 2 described in the text. Error bars correspond to the 95\% confidence bounds. The Curie-Weiss fit is shown by solid line (fitted value of CW tempearture is $T_0=436 \pm 36$ K. Dashed line shows CW fit with $T_0$ fixed at value 378 K. The inset shows the temperature dependeces for the relative rates of growth corresponding to the curves in the main plot.
}
\end{figure}

Fig. \ref{fig_cp_cw} shows the temperature dependence of the central peak intensity normalized to the absolute value of the square of the TO phonon structure factor, which is assumed to be temperature independent. The solid line shows the least-squares fit by Eq. \eqref{eq_icp_t}. The fit gives a value of $T^\text{CP}_0=436 \pm 36$ K. This value is numerically similar to $T_0^\text{Bussmann}$, but is substantially larger than  $T_0^\text{Samara}$. The data was also fitted by a CW law with $T_0$ fixed at $T_0^\text{Samara}$; the result is shown by the dashed line. An additional fit by a formula equivalent to \eqref{eq_icp_t}, but with the denominator $(T-T_0^\text{CP})$ replaced by $(T-T_0^\text{CP})^n$, where $n$ is an additional free parameter, gives $T_0=285$ K and $n=2.4$. This fit (dotted line) appears slightly more consistent, but also not completely adequate.  None of these fits is really satisfactory. 

The inset of Fig. \ref{fig_cp_cw} shows the relative rates of growth (defined by Eq. \eqref{eq_rrog}), by the same line types, as in the main plot. Near the transition temperature (about 470 K) $R(T)$ for the susceptibility associated with the central peak (either dotted or solid lines) is larger than the one for $\epsilon(T)$ (dashed line).

\subsubsection{Scenario 3. Both the TA phonon and the central peak are temperature-dependent.}

\begin{figure}
\includegraphics[width=.99\columnwidth, clip=true, trim= 10 0 10 0]{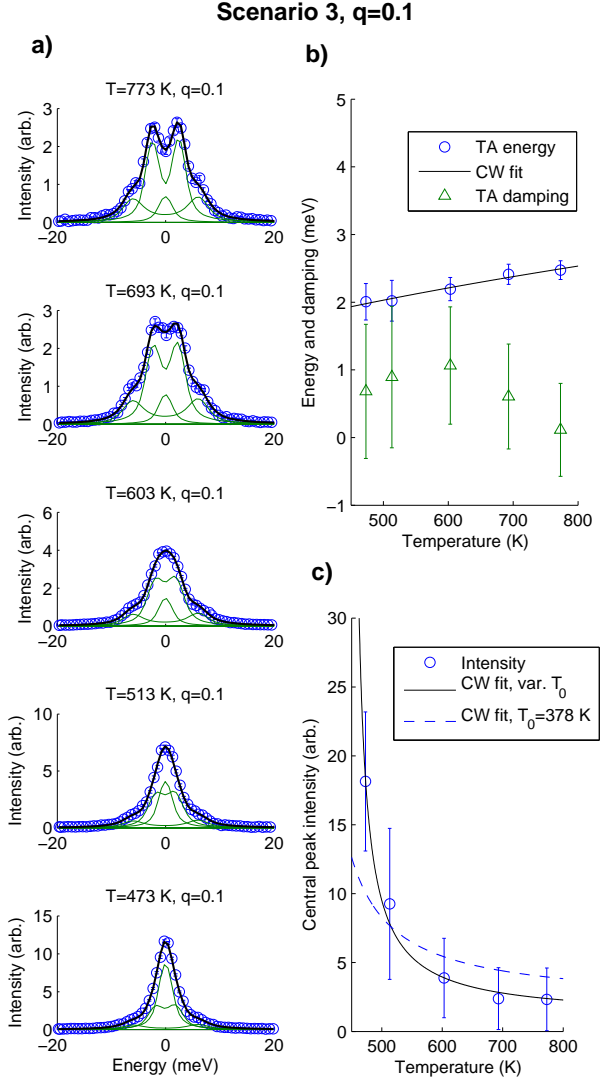}
\caption{\label{fig_everything}
Fit results within Scenario 3 for data at $\vec{Q}=(2.1, 0.1, 0)$. (a) Temperature evolution of the spectra and the corresponding fits. Error bars correspond to 95\% confidence bounds, they are smaller than symbols. (b) Temperature dependence of the TA phonon energy (circles) and damping constant (triangles). The Curie-Weiss fit of the energy is represented by the solid line. (c) Temperature dependence of the central peak intensity and the corresponding CW fits with variable $T_0$ (solid line) and $T_0=T^\text{Dielectric}_0=378$ K (dashed line).
}
\end{figure}

This scenario is a hybrid of Scenario 1 and Scenario 2. The temperature dependence is admitted in the TA phonon energy and damping and in the magnitude of the central peak. The constraints are the same as in the Scenario 1. The results of the fit are shown in Fig. \ref{fig_everything}a. The temperature dependence of the TA energy and damping are weak (Fig. \ref{fig_everything}b): the Curie-Weiss fit to the temperature dependence of the energy gives $T_0<0$ K.  The central peak intensity is treated in the same manner as in Scenario 2 (Fig. \ref{fig_everything}c) resulting in $T_0=447 \pm 9.5$ K. Apparently, the less constrained Scenario 3 gives qualitatively the same results as the more constrained Scenario 2. The temperature dependence of the intensity is more consistent with the CW law than within Scenario 2. On the other hand, the relative rate of growth is, again, larger than the one for $\epsilon(T)$.

\subsubsection{Comparison of the fit statistics for q=0.1}

From the visual inspection of Fig. \ref{fig_q01_comparison} it is evident that Scenarios 1 and 2 produce equivalently convincing fits. Apart from the visual inspection, a common way of evaluating the ``fit quality'' is by comparing the value of the root mean square error for different models, defined as \cite{Numerical}
\begin{equation}\label{eq_rmse}
R=  \frac{  
\sqrt{  \sum_{i=1}^N \frac{(I^\text{Exp}_i - I^\text{Model}_i)^2 }{\sigma^2_i}  } 
}{N-M},
\end{equation}
where $I^\text{Exp}_i$ -- the data points, $I^\text{Model}_i$ -- corresponding model calculations, $\sigma_i$ -- expected variances for data points (size of the error bars), $N$ -- number of data points, $M$ -- number of variable parameters. It is the same parameter which is minimized during the fit process. In Scenario 1 the value of $R_\text{S1} = 1.4$, in Scenario 2 $R_\text{S2} =1.44$. Consequently, this estimate of the ``fit quality'' assigns only a minor favor to Scenario 1 over Scenario 2.
Scenario 3 has more independent variables than Scenarios 1 and 2; in fact the models of Scenarios 1 and 2 are the limiting cases of the model in Scenario 3. The aim of Scenario 3 can therefore be, at the best, to examine what the data fitting algorithm will ``prefer'' when provided with increased flexibility. Apparently it ``prefers'' Scenario 2. However, one shall expect it to ``prefer'' as well Scenario 1 when the starting point in the data fitting algorithm would be chosen accordingly, because the corresponding local minimum in the objective function is deeper for Scenario 1 ($R_\text{S1} < R_\text{S2}$). 

The conclusion of this section is that the temperature dependence of the IXS spectra at $q=0.1$ can be described equally well by the two different models (Scenarios 1 and 2) that allow a reasonable conditionality of the optimization problem. The ``physical relevance'' of these two models shall be assessed at a later stage, in connection with the results obtained at different wavevector transfers.

\subsection{Modeling the temperature dependence of the signal in the remaining part of $\Gamma$-M direction}

When dealing with the data recorded at $\vec{Q}=(2.1,0.1,0)$ we tried the three scenarios. Although Scenarios 1 and 2 have different number of independent fit parameters, they both cannot be further reduced by imposing more constraints without losing the ability to give a ``good fit''. Scenario 3, on the other hand, is over-parametrized which makes, in general, the fitting problem less well conditioned. Nevertheless, it provided reasonable results for the IXS data at $q=0.1$. The spectra at larger $q$ are in general less informative, meaning that the peaks are less well resolved. This makes Scenario 3 practically inapplicable -- the fitting problem becomes ill-conditioned and gives irregular results. For the other $q$-points along the $\Gamma$-M direction we therefore used only the first two scenarios. 
The results of the fits are shown in Appendix \ref{app_gm}.

Generally, at larger $q$ the TO phonon resonance is less resolved than at $q=0.1$. For $q=0.3$ and larger wavevectors it becomes overdamped. The visual appearance of the fits, as in the case of $q=0.1$ does not allow favoring one scenario over the other. The result of the numerical evaluation using \eqref{eq_rmse} is shown in Table \ref{tab_gm} for all the points along the $\Gamma$-M direction. In all cases the root mean square error is slightly less for Scenario 1. We do not claim, however, that this difference can be used to decide which scenario is more likely to be correct. 

We attempted to bring more insight into the problem of distinguishing the two scenarios by considering the temperature evolution of the spectra at wavevectors, which are slighlyt away from the $\Gamma$-M direction. In particular, we considered the point $\vec{q}=(0.12,0.21,0)$, which is, on one hand, close enough to the $\Gamma$-M direction to be influenced by the same physical effects and, on the other hand, is far enough to have larger energy of the TA phonon to provide a better distinction of its temperature dependence. This analysis was, however, inconclusive.

\begin{table}
\begin{tabular}{l r*{5}{r}  } 
    $q$ & 0.1 & 0.2 & 0.3 & 0.4 & 0.5 \\ \hline
	Scenario 1 & 1.40 & 1.23 & 1.34 & 1.29 & 1.27 \\
	Scenario 2 & 1.44 & 1.38 & 1.64 & 1.48 & 1.36 \\
\end{tabular}
\caption{\label{tab_gm}
The root mean square errors characterizing the fits of the data along the $\Gamma$-M direction using the two fitting scenarios, as described in the text.
}
\end{table}

\begin{figure}
\includegraphics[width=.7\columnwidth, clip=false, trim= 10 0 10 0]{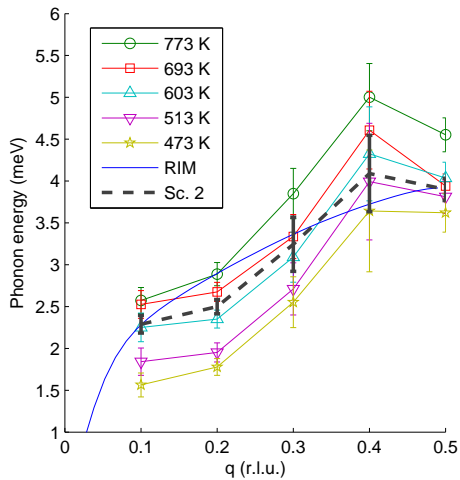}
\caption{\label{fig_fspb}
Temperature dependence of the TA phonon energy along the $\Gamma$-M direction, as obtained by fits according to Scenario 1. Solid lines connecting symbols are guides to the eye. Smooth symbol-free solid line -- calculation by the rigid ion model from Fig. \ref{fig_ixs}. The dashed line shows the TA phonon dispersion, as determined according to the Scenario 2.
}
\end{figure}

The temperature dependence of the TA phonon dispersion curve along the $\Gamma$-M direction, as determined according to the Scenario 1, is shown in Fig. \ref{fig_fspb}. The energy decreases monotonically for $q\geq 0.1$ at all temperatures. At large $q$, the fit with Scenario 1 gives slightly modified energy for the phonons at T=773 K, as compared to the energy determined by independent fitting of just the high-temperature data. The independently determined high-temperature dispersion curve from Fig. \ref{fig_ixs} is shown in Fig. \ref{fig_fspb} by a solid line. On the other hand, at $q\leq 0.2$ Scenario 1 is consistent with the result of the independent fit. For Scenario 2 the situation is different: at $q \geq 0.3$ the solid curve (independent high-temperature fit) is within the error bars of the points obtained within Scenario 2, while for $q = 0.2$ the curve is substantially higher than the point from Scenario 2.  

\subsection{R-point}

\begin{figure}
\includegraphics[width=.99\columnwidth, clip=false, trim= 10 0 10 0]{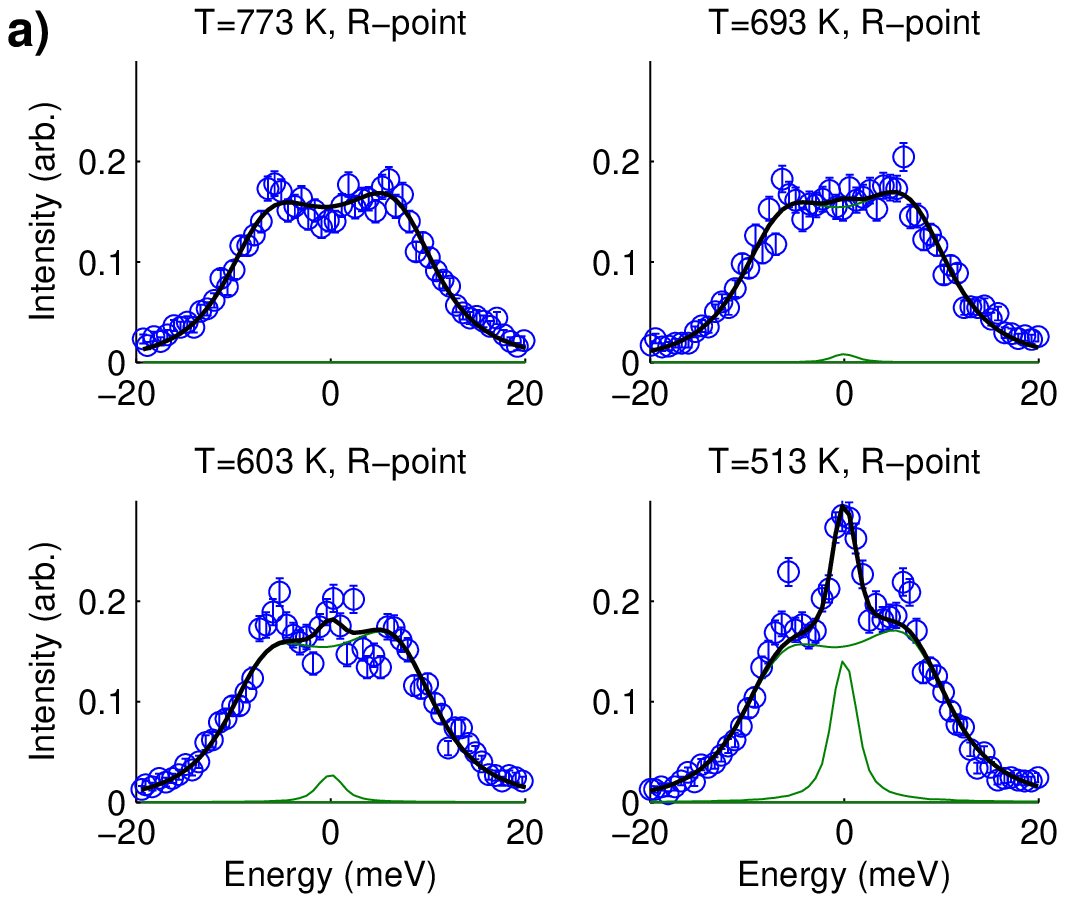}
\includegraphics[width=.7\columnwidth, clip=false, trim= 10 0 10 0]{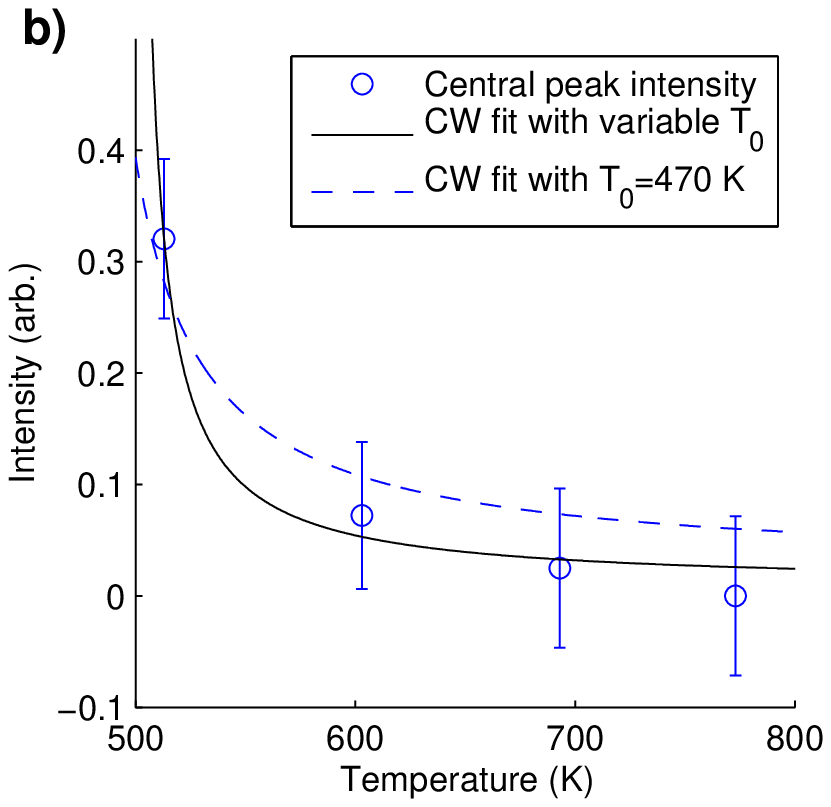}
\caption{\label{fig_r}
(a) Temperature dependence of the IXS spectra, obtained at the R-point ($\vec{Q}=(3.5,2.5,0.5)$), and the results of the corresponding fit.
(b) Temperature dependence of the central peak intensity and the Curie-Weiss fit with variable $T_0$ (solid line) and with $T_0$ fixed at 470 K (transition temperature) (dashed line). With variable CW temperature the fit gives $T_0=500$ K, which is larger than the temperature of the transition.
}
\end{figure}

The spectra for the R-point ($\vec{Q}=(3.5,2.5,0.5)$) are shown in Fig. \ref{fig_r}a. At this point we took the spectra at the same temperatures as the spectra for the $\Gamma$-M direction, except the lowest temperature (473 K). The spectra contain the Stokes and anti-Stokes resonances of one phonon and the central peak. 

The independent fits for these spectra admit a weak softening of the phonon resonance -- about 1 meV on cooling from 773 K to 513 K. The energy and damping constant of the phonon in this region are estimated to be 9 and 10 meV, respectively. The admitted level of softening, relative to the average energy (9 meV) is comparable to the one suggested by the analysis within Scenario 3 for the TA phonon at $\vec{q}=(0.1, 0.1, 0)$ (see Fig. \ref{fig_everything}b). Neglecting this possible weak softening, we performed the fit of the spectra at R-point similarly to the fits in Scenario 2 for the $\Gamma$-M direction, assuming the energy and damping constant of the phonon to be temperature-independent, the result is in Fig.\ref{fig_r}a. The temperature dependence of the central peak intensity (normalized by the squared modulus of the phonon structure factor) is shown in Fig. \ref{fig_r}b. 
We provide two CW fits of the intensity. The first fit (solid line), with variable CW temperature, yields $T_0=500 \pm 25$ K . This temperature is above the transition temperature. For comparison we provide a fit with $T_0$ fixed at the transition temperature (dashed line). 
The temperature dependence of the central peak intensity at the $R$-point is inconsistent with the CW law with critical temperature below the transition temperature. This renders the assumption that the central peak at the $R$-point can be directly associated with a susceptibility following CW law inconsistent. A more complex model is required for the consistent description of the data. 

\section{Summary of observations} \label{sec_summary}

In the previous section we gave thoroughly illustrated results of the IXS measurements and of their basic analysis. This analysis was limited to (a) fitting the IXS spectra by model functions, (b) fitting the dispersion curves by the rigid ion model and (c) fitting the temperature dependences by the Curie-Weiss law in an appropriate form. Here we summarize the findings.

\textbf{1. $\Gamma$-M direction, transverse displacements}
  \begin{enumerate}
    \item At high temperatures, there is a relatively flat TA branch, with a sharp change in the phonon group velocity at $q=q_\text{AC}$. At this point there is an avoided-crossing between TA and TO branches.
    \item The TO branch, at $q>q_\text{AC}$, is temperature-independent, at least within the accuracy of our data.
    \item On cooling down, there is an intensity increase of small-energy-transfer-scattering (SETS). By SETS we mean the scattering due to everything except the TO phonons.
    \item The data admits different models for the SETS growth.
    \item In Scenario 1 the SETS growth is due to the TA branch softening. The branch softens as a whole for $q>q_\text{AC}$. The temperature dependences of the susceptibility at $q=0$ (ferroelectric soft mode) and at $q=0.1$ (TA mode just after avoided-crossing) can be described by the Curie-Weiss law with similar critical temperatures.
    \item In Scenario 2 the SETS growth is due to the emerging central peak. Both TA and TO phonons are unaltered. The relative rate of growth for the susceptibility, straightforwardly extracted from the central peak intensity at $q=0.1$, is larger than the one for $\epsilon(T)$.
    \item Scenarios 1 and 2 are minimal in the sense that additional constraints on the fitting parameters do not yield plausible fits. The current data does not admit a reliable analysis by a more complex model, such as Scenario 3.
  \end{enumerate}

\textbf{2. $\Gamma$-X direction, transverse displacements}

  	No temperature dependence, no central peak. Exception -- the proximity of the $\Gamma$ point.

\textbf{3. R-point}

  	Nearly temperature-independent phonon resonance, growing central peak. The relative rate of growth, deduced from the central peak temperature dependence, is too high to be compatible with the CW law with critical temperature below the transition temperature.

\section{Interpretation and discussion} \label{sec_discussion}

The aim of the current discussion is mainly limited to addressing the question 
``What kind of phase transition can be compatible with the properties of the lattice dynamics in the paraelectric phase?''.
The non-Bragg scattering experiment in the high-symmetry phase can only give the information on the susceptibility of the crystal with respect to particular patterns of ionic displacements, as defined by Eq. \eqref{eq_chi_jj}. So we can reformulate the main question as
``Which trends in the observed temperature dependences of different sorts of susceptibility can suggest particular static displacement patterns in the low-symmetry phase?''.

\subsection{Ferroelectricity}

There is no unambiguous evidence whether the corresponding $q=0$ soft TO mode should be an underdamped phonon or a soft mode of relaxational character. 
In the case of PbZrO$_3$, which can be expected to have dynamics similar to PbHfO$_3$, the infra-red studies show a heavily damped response due to the soft mode \cite{Ostapchuk2001}, supporting rather the latter option. Further support comes from the results of neutron pair distribution function analysis in the high-temperature phase of PbZrO$_3$ that indicate the presence of quasistatic displacements \cite{Teslic1998}. For PbHfO$_3$ there are similar, although less detailed evidences \cite{Kwapulinski1994}. The study of the dynamics at the zone center was carried out by Raman scattering in Ref. \onlinecite{Jankowska1995}. A similar growth of the small energy transfer signal for the zone center was observed as in the present work for finite wavevectors. The authors of Ref. \onlinecite{Jankowska1995} discuss the question of displacive versus order-disorder character of the ferroelectric soft mode, but do not give an unambiguous conclusion.

\subsection{Antiferroelectric and incommensurate modulations}

\subsubsection{Displacive-like transition} \label{subsubsec_dlt}
The simplest possible interpretation of the IXS spectra in the $\Gamma$-M direction is the following. We admit that the model in Scenario 1 is the correct one. This means that all the temperature evolution can be explained in the soft phonon language. 
In this case the flat part of the TA branch for $q>q_\text{AC}$ is temperature dependent and the rate of softening, $d\omega_\text{TA}(\vec{q})/dT$, is similar in magnitude for all considered $q$-points. The data do not reveal any minimum of the TA dispersion curve. This is in direct analogy with the TA branch in PbZrO$_3$ \cite{Tagantsev2013,Burkovsky2014}. Following Ref. \onlinecite{Tagantsev2013} one may suggest that the formation of the commensurate antiferroelectric modulation occurs due to the first-order phase transition conditioned upon Umklapp interactions. On the other hand, the incommensurate phase transition can also be possible due to the first order transition. The absence of a minimum in the dispersion curve, in this context, only shows that the transition should be not very similar to the second order phase transition. An adequate estimate of this ``dissimilarity'' could be the difference between the critical temperature $T^\text{TA}_0 \approx 295 K$ and the actual transition temperature. This difference is about 170 K. So, the actual transition can be ``sufficiently dissimilar'' with the second order phase transition, despite the observed mode softening. 

\subsubsection{Order-disorder-like transition} 
An alternative interpretation arises when we assume that Scenario 2 is correct.
The susceptibility $\chi^\text{Stat}_\text{CP}(\vec{q})$, linked with the central peak intensity via \eqref{eq_sc_fun_int}, increases on cooling for the whole $\Gamma$-M direction, so one may expect a first-order antiferroelectric or incommensurate transition with a modulation wavevector along this direction.
The phonons at $q>q_\text{AC}$ are assumed to be temperature independent. The quasiharmonic models, such as the ``polarizability model'' from Ref. \cite{Bussmann2014}, or the rigid ion model, as employed by us, suggest that the zone center TO phonon shall be also temperature-independent in this case. 
So, the temperature dependences of both the dielectric ($q=0$) and the generally  incommensurate ($q>0$) susceptibilities should be in some way determined by a temperature dependence of $\chi^\text{Stat}_\text{CP}(\vec{q})$.

First we consider the case in which both these susceptibilities are just proportional to $\chi^\text{Stat}_\text{CP}(\vec{q})$ at corresponding wavevectors. 
With this assumption it turns out (as was already noted in the section dealing with the data treatment according to Scenario 2), that the relative rate of the susceptibility increase with respect to modulations with $q=0.1$ is larger than the one for $\epsilon(T)$, at least for $T_c < T <T_c+\Delta T$. On the other hand, $\chi^\text{Stat}_\text{CP}(\vec{q})$ for $q=0.1$ does not appear to increase slower than the one for larger $q$ (as determined in Scenario 2, not shown). So, within the present assumption, the relative growth rate for the susceptibility at $q=0.1$  is larger than the one for $q=0$ and $q>0.1$. This is unlikely. In the presence of the ferroelectric soft mode, $\chi^\text{Stat}_\text{CP}(\vec{q})$ should be a function, decreasing with $q$ increase \cite{Blinc1974}. The assumption that the susceptibility is just proportional to $\chi^\text{Stat}_\text{CP}(\vec{q})$ with respect to modulations is therefore unlikely.


In an alternative scenario we admit that the TA phonon and the central peak are due to displacements with similar patterns. This assumption corroborates well with the fact that the central peak is observed in the same areas of the reciprocal space, where the TA branch is unusually flat.
In this case it is natural to assume that at $q=0.1$ and for larger values of $q$ the susceptibility with respect to the formation of the modulation waves is jointly reflected by the susceptibility with respect to the displacements in the TA phonon and by the susceptibility due to the redistribution of the quasistatic displacements. In analogy with that, the dielectric response ($q=0$) is the result of both the ionic shifts in the TO phonon mode and the redistribution of the quasi-static displacements (that would be seen as a central peak in a scattering experiment). The simplest case assumes that the patterns of ionic displacements in the TA branch and in the quasistatic disorder are the same. This implies equal structure factors for these two components. The integral intensity over SETS is thus determined by the sum $\chi^\text{Static}_\text{CP}(\vec{q}) + \chi^\text{Static}_\text{TA}(\vec{q})=\chi^\text{Static}_\text{SETS}(\vec{q})$ as (from Eq. \eqref{eq_sc_fun_int})
\begin{equation}\label{eq_sets}
S^\text{Int}_\text{SETS}(\vec{q}) = \frac{|G_j (\vec{Q},\vec{q}) |^2}{\beta} \chi^\text{Static}_\text{SETS}(\vec{q}).
\end{equation}
The temperature dependence of $\chi^\text{Static}_\text{SETS}(\vec{q})$ is nearly the same as the one, determined for $\chi^\text{Static}_\text{TA}(\vec{q})$ in Scenario 1, because the parameters of the TO phonon in both scenarios are nearly the same implying the same integrals over SETS. In this case the consideration of the temperature dependence of the static susceptibility is completely equivalent to the one for the displacive-like scenario.

\subsubsection{Generalized viewpoint}

In the preceding paragraphs we considered the two viewpoints on the pre-transitional dynamics. In the former, the crystal approaches the displacive-like transition driven by the soft phonon. In the latter, it approaches the order-disorder-like transition, where the susceptibility increase is reflected in the spectra by an increase of the displacement-displacement correlation function (central peak) for the quasistatic displacements. These two viewpoints are the limiting cases of the following more general consideration. The dynamic scattering function for the lattice mode $j$ is written as  \cite{Bruce1981}
\begin{equation}
S_j(\vec{Q},\omega)=  |G_j(\vec{Q,q})|^2 \frac{1}{\pi \beta \omega} \chi_j^\text{I}(\vec{q},\omega),
\end{equation}
where $G_j(\vec{Q,q})$ -- structure factor for the lattice mode and $\chi_j^\text{I}(\vec{q},\omega)$ -- imaginary part of the dynamic susceptibility. In the case of the displacive-like transition viewpoint the dynamic susceptibility is the one of the damped harmonic oscillator $\chi_\text{DHO}(\omega,\omega_j,\Gamma) = (\omega_j^2-\omega^2-i\omega\Gamma)^{-1}$. In the case of the order-disorder-like transition, the dynamic susceptibility can be represented by $\chi_\text{OD+DHO} = \chi_\text{DHO}(\omega,\omega_\text{CP},\Gamma^\text{BIG})+ \chi_\text{DHO}(\omega,\omega_\text{TA},\Gamma)$, where the part responsible for the central peak is parametrized by a DHO susceptibility function with very large damping constant $\Gamma^\text{BIG}$. The contributions of the two terms to the cumulative static susceptibility $\chi^\text{Static}_\text{SETS}$ are given by the parameters $\omega_\text{CP}$ and $\omega_\text{TA}$ via Eq. \eqref{eq_chi_omega}. Among the possible more sophisticated models for the dynamic susceptibility function we shall mention the one introduced in Refs. \onlinecite{Shapiro1972,Axe1973} and used in Ref. \onlinecite{Tagantsev2013,Burkovsky2014} for the interpretation of the IXS spectra in PbZrO$_3$. In the current paper we do not attempt using it for PbHfO$_3$ because the IXS spectra do not contain enough details to make its application reliable for practical data analysis.

\subsubsection{Summary on the $\Gamma$-M direction}

From the considerations above it turns out that it is unlikely that we shall associate the central peak in Scenario 2 with the static susceptibility and that the integral scattering intensity (SETS) should be used for that purpose instead. The static susceptibility $\chi^\text{Static}_\text{SETS}(\vec{q})$, for small $q$, appears to have a similar temperature dependence as the one of $\epsilon(T)$. Consequently, the transition to the modulated IC phase can be as anticipated from the IXS results, as ferroelectricity can be anticipated from the dielectric results. 

Regarding the nature of the dynamic susceptibility, no unambiguous conclusion can currently been drawn. Although the analysis of spectra at $\vec{q}=(0.12,0.21,0)$ suggests that the TA phonon has a temperature dependence, the magnitude of this temperature dependence cannot be traced quantitatively and the presence of the central peak cannot be ruled out. The presence of the central peak is natural to expect due to the known evidences in favor of the presence of quasistatic displacements in the paraelectric phase \cite{Kwapulinski1994}. 
We note that the problem of theoretical understanding of the difference between displacive and order-disorder regimes of the phase transition, and possible crossovers between them, is quite challenging, and was intensively studied in the context of simple models for the Hamiltonian, such as the so-called $\phi^4$ model \cite{Schneider1976,Bruce1981}. The relationship between soft modes and central peaks is also an intensively studied topic \cite{Fleury1982}. We think that there is space for a deeper understanding of the relationship between the rather broad spectrum of theoretical results and our experimental observations.

\subsection{R-point order parameter}

At the R-point we observed nearly temperature-independent phonons and a strongly temperature dependent central peak. In the absence of conclusive information on the displacement patterns in these two spectral features we can only sketch the possible variants of the interpretation. In the antiferroelectric phase of PbHfO$_3$ the superstructure reflections at the R-point (in pseudocubic reference frame) are associated with oxygen octahedra rotations. We suggest that the growth of the central peak on cooling in the paraelectric phase reflects the quasi-static disorder in the shifts of oxygen atoms. As it was shown in the corresponding data analysis section, the temperature dependence of the central peak intensity is unlikely to be consistent with the CW relationship with $T_0$ below the transition temperature: the relative rate of growth for the corresponding susceptibility is too large near the transition. On the other hand, it might be possible that this temperature dependence can be consistently described by more sophisticated models such as in Ref. \onlinecite{Shapiro1972}.

\subsection{Comparison with PbZrO$_3$}

The current study reveals both similarities and potential differences between the characters of the lattice dynamics in PbHfO$_3$ and PbZrO$_3$ \cite{Tagantsev2013,Burkovsky2014}. In both crystals there is an increase of the static susceptibility with respect to the transverse modulations with the wavevector in the $\Gamma$-M direction. The measurements in Refs. \onlinecite{Tagantsev2013,Burkovsky2014} and in the current study were done in different Brillouin zones (BZs). In PbZrO$_3$ (measurements in the BZ (3 2 0)) the central peak was reported for all temperatures up to $T=780$ K. In PbHfO$_3$ (BZ (2 0 0)) there is no central peak at high temperatures ($T=773$ K) and it cannot be unambiguously identified at lower temperatures: the fits with and without the central peak describe the spectra equally well for $q$-points exactly at the $\Gamma$-M direction. On the other hand, in PbHfO$_3$ we recorded the well-observed, underdamped up to $q\approx 0.2$, TO phonon branch which apparently has an avoided-crossing with the TA branch and is the continuation of the initial transverse acoustic slope at $q>q_\text{AC}$. The notable difference is in the temperature dependence of the R-point spectra. For PbZrO$_3$ the R-point central peak was reported to be weakly temperature dependent, while in PbHfO$_3$ it is strongly temperature dependent near the transition temperature.

\section{Conclusion} \label{sec_conclusion}

We reported a comprehensive analysis of the inelastic X-ray scattering data in PbHfO$_3$. The  temperature dependence of the small energy transfer scattering points towards the possibility of formation of incommensurate or antiferroelectric phases in the course of the first-order phase transition. Generally it supports the result of Refs. \onlinecite{Tagantsev2013,Burkovsky2014} that antiferroelectricity can be a lock-in phase in the incommensurate phase transition scenario. One might expect the incommensurate phase to occur in the temperature range currently assigned to intermediate AFE phase(s) in PbHfO$_3$, which corroborates with suggestion of Ref. \onlinecite{Bussmann2015}, based on the analysis of birefringence measurements.

It turns out that the zone-center ferroelectric soft mode (of whatever displacive or order-disorder origin) and the generally  incommensurate soft mode on the right of the avoided-crossing region are likely to correspond to similar ionic displacements that occur homogeneously for the zone-center soft mode and in the form of modulations for the generally  incommensurate soft mode. This assumption is supported by the fact that the temperature dependence of the susceptibility on the left and on the right hand sides of the avoided-crossing region is approximately the same.

The temperature evolution of the IXS spectra in PbHfO$_3$, similar to the case of PbZrO$_3$, is quite challenging for an unambiguous analysis. In this paper we rigorously explored up to which extent one may increase the flexibility of the data fitting model without compromising the conditionality of the data fitting problem. With the current energy resolution (3 meV), the problem of distinguishing the correct variant of the dynamic susceptibility function appears ill-defined, while the static susceptibility can be well analyzed. One may expect that an improved energy resolution can help settling the question of the dynamic susceptibility function in PbHfO$_3$. In this context PbHfO$_3$ can be more prospective than PbZrO$_3$. The spectra obtained in this study for PbHfO$_3$ appear, on visual inspection, more detailed than the ones from Refs. \onlinecite{Tagantsev2013,Burkovsky2014}, taken with a better resolution of 1.5 meV. In particular the TA phonon peaks are better resolved in the current study at high temperatures.

\begin{acknowledgements}
D. Chernyshov is acknowledged for useful discussions. A. Bosak is acknowldeged for his help with the sample preparation, useful discussions and suggestions.
\end{acknowledgements}

\appendix
\section{Spectra for the $\Gamma$-M direction} \label{app_gm}

\begin{figure*}
\includegraphics[width=\textwidth, clip=false, trim= 20 0 20 0]{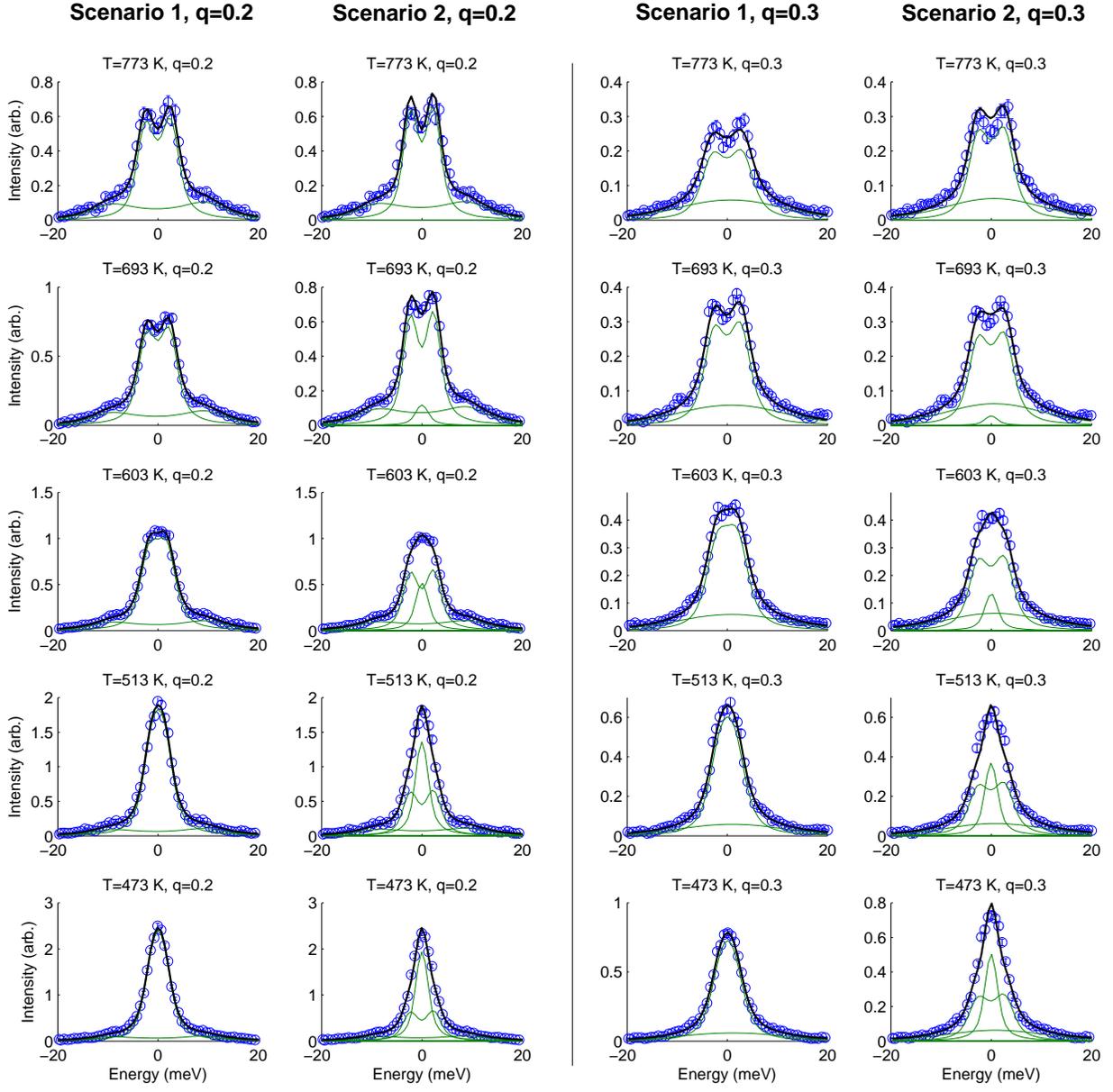}
\caption{\label{fig_spectra_23}
IXS spectra of PbHfO$_3$ corresponding to $\vec{Q}=(2.2,0.2,0)$ and $\vec{Q}=(2.3,0.3,0)$ treated according to Scenarios 1 and 2. 
}
\end{figure*}

\begin{figure*}
\includegraphics[width=\textwidth, clip=false, trim= 20 0 20 0]{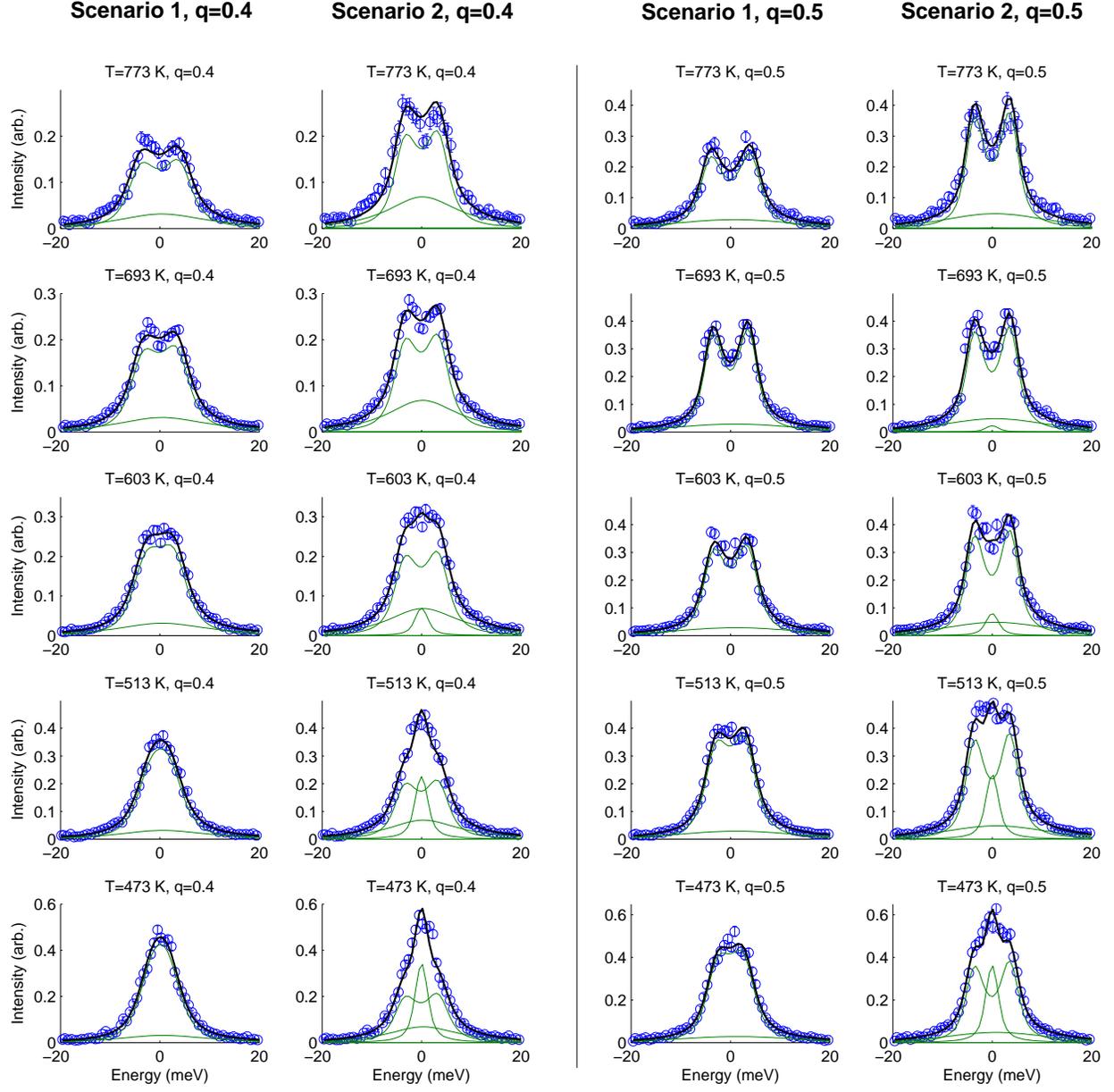}
\caption{\label{fig_spectra_45}
The IXS spectra corresponding to $\vec{Q}=(2.4,0.4,0)$ and $\vec{Q}=(2.5,0.5,0)$ treated according to the Scenarios 1 and 2. 
}
\end{figure*}

The results for the fits of the data along the $\Gamma$-M direction are shown in Fig. \ref{fig_spectra_23} (for $q=0.2$ and $q=0.3$) and in Fig. \ref{fig_spectra_45} (for $q=0.4$ and $q=0.5$). The presentation method is the same as in Fig. \ref{fig_q01_comparison} -- for each $q$ there are two adjacent panels showing the fits according to the two scenarios. 
Starting from $q=0.3$ the signal from the TO phonon becomes poorly resolved and appears as an overdamped DHO signal in the fits. The description of the shape is not always satisfactory in this region. In particular, for $q=0.3$ the shape of the experimental curve does not strictly follow the fit at energy transfers around 7-8 meV. The current data quality does not allow a better description.

\bibliographystyle{apsrev4-1}
\bibliography{bibliography_PHO}

\end{document}